# Terahertz frequency combs exploiting an on-chip solution processed graphene-quantum cascade laser coupled-cavity architecture

Francesco P. Mezzapesa,[1] Katia Garrasi,[1] Johannes Schmidt,[1] Luca Salemi,[1] Valentino Pistore,[1] Lianhe Li,[2] A. Giles Davies,[2] Edmund H. Linfield,[2] Michael Riesch,[3] Christian Jirauschek,[3] Tian Carey,[4] Felice Torrisi,[4+] Andrea C. Ferrari,[4] and Miriam S. Vitiello[1*]

[1] *NEST, CNR - Istituto Nanoscienze and Scuola Normale Superiore, Piazza San Silvestro 12, 56127, Pisa, Italy*
[2] *School of Electronic and Electrical Engineering, University of Leeds, Leeds LS2 9JT, UK*
[3] *Department of Electrical and Computer Engineering, Technical University of Munich, Arcisstr. 21, 80333 Munich, DE*
[4] *Cambridge Graphene Centre, University of Cambridge, Cambridge, CB3 0FA, UK*
[+] *present address: Molecular Science Research Hub, Imperial College London, London W12 0BZ, UK*

[*]miriam.vitiello@sns.it

**The ability to engineer quantum-cascade-lasers (QCLs) with ultrabroad gain spectra and with a full compensation of the group velocity dispersion, at Terahertz (THz) frequencies, is a fundamental need for devising monolithic and miniaturized optical frequency-comb-synthesizers (FCS) in the far-infrared. In a THz QCL four-wave mixing, driven by the intrinsic third-order susceptibility of the intersubband gain medium, self-lock the optical modes in phase, allowing stable comb operation, albeit over a restricted dynamic range (~ 20% of the laser operational range). Here, we engineer miniaturized THz FCSs comprising a heterogeneous THz QCL integrated with a tightly-coupled on-chip solution-processed graphene saturable-absorber reflector that preserves phase-coherence between lasing modes even when four-wave mixing no longer provides dispersion compensation. This enables a high-power (8 mW) FCS with over 90 optical modes to be demonstrated, over more than 55% of the laser operational range. Furthermore, stable injection-locking is showed, paving the way to impact a number of key applications, including high-precision tuneable broadband-spectroscopy and quantum-metrology.**

Optical frequency comb synthesizers (FCSs) enable broadband coherent light sources to be developed that consist of a large number of equally spaced lasing modes.[1] Chip-scale, broadband, monolithic, high brightness FCS sources at THz frequencies are needed for metrology,[2-6] ultra-high speed communications,[6] coherent nano-tomography,[7] near-field broadband nanoscopy and for opening new avenues in high-resolution broadband molecular spectroscopy, manipulation of complex molecules and cold atoms, astronomy, and attosecond science.[8]

QCLs are the highest brightness miniaturized sources in the infrared (IR).[9,10] They combine an inherently high spectral purity,[11-12] with a very broad bandwidth,[13-15] Watt-level output-powers,[16] and a long upper-state lifetime (10 ps). Broad emission in the THz range can be attained by precise quantum-tailoring the gain medium to host heterogeneous stacks of individual active regions, which themselves are incorporated into a monolithic microstrip-line metal-metal resonator.[17,18] The cascading design of individual emitters at complementary wavelengths creates a flat broad-gain at a desired bias point, which is usually slightly above the onset of multimode emission. In addition, the long upper state lifetime ($\geq$5-10 ps) [9,10,19] of THz QCLs inhibits mode-locking but favours phase-matching between the cavity modes driven by the ultrafast non-linearity (four-wave mixing (FWM)) spontaneously arising in the intersubband gain medium.[13,15,20] The resulting stable frequency comb regime tends, however, to be restricted to lower injection currents, close to the onset of multimode emission, owing to the intersubband bias-dependent contribution to the group velocity dispersion. [13,15,20] The heterogeneous nature of the gain media then entangles the dispersion dynamics at other biases.[20] Tailoring the dynamic (bias-dependent) contribution to the chromatic dispersion is, therefore, crucial for establishing perfectly spaced, phase-locked, high-intensity modes spanning the entire dynamic range of the laser.

The use of chirped mirrors[20] and/or coupled cavities[21-23] have been proposed for GVD compensation in homogeneous[20,21] and heterogeneous[22,23] THz QCL FCSs, respectively. In the first case, the corrugation length and the tapering period of a chirped cavity was designed to optimize the anomalous group delay dispersion (GDD) in a specific, narrow, range of biases, leading to an FCS extending over 24% of the laser dynamic range, covering a 0.6 THz spectral bandwidth, but with an uneven distribution of power amongst the 60 modes.[20] In the second case, a coupled cavity, mimicking a Gires–Tournois interferometer (GTI),[21] was monolithically integrated in front of the THz QCL cavity, introducing chromatic dispersion that compensates for that in the gain medium.[21] When DC biased, the small coupled-cavity section gave rise to an FCS over the whole QCL dynamic range, although there was only a very limited amount of optical power (µW) over a < 0.4 THz bandwidth, distributed amongst a few, irregularly distributed modes of dissimilar intensities; this was a result of dual-cavity induced suppression of the multi-mode operation regime.[21] Such coupled cavity architectures do, however, prove to be extremely beneficial for generating ultrashort THz pulses through injection seeding.[22] Using a very small intra-cavity spacing (1.5-2µm), and predefined coupled-cavity lengths, effective chromatic dispersion compensation was achieved. However, this was only possible over a very limited frequency range, due to the nonlinear phase-frequency relation associated with the dispersion. Indeed, oscillation in the group delay dispersion (GDD) was expected, as the phase of the reflected light, and the GDD, change periodically with the optical frequency owing to resonance effects.[22]

Despite these advances, there remains a lack of miniaturized technologies for high-power (> 5 mW), broadband (~ 1 THz bandwidth) THz QCL FCSs, providing phase-locked, evenly spaced, lines of comparable intensity over the entire laser dynamic range. Furthermore, for high-precision ($10^{-11}$) and high-sensitivity (part in $10^{-6}$ cm$^{-1}$/Hz$^{-1/2}$) metrological applications, there is a further requirement for fine frequency tunability of the comb lines over a broad spectral window, combined with an (ideally) zero time jitter of the phase-locked modes and small phase fluctuations; this would allow THz QCL FCSs to be fully stabilized against primary frequency standards.[24]

In this work, we present a record dynamic range THz QCL FCS based on a monolithic coupled-cavity architecture comprising a heterogeneous THz-frequency QCL, with a wide (3.2) dynamic range, and an on-chip solution-processed multilayer saturable absorber graphene (GSA) reflector, and demonstrate stable injection locking. The gapless nature of the reflector, and the related frequency-independent absorption, ultrafast recovery time,[25] low saturation fluence,[26] and ease of fabrication[27] and integration,[28,29] makes graphene an appealing non-linear optical component in the infrared, and ideal for developing THz QCL FCSs, with performances far-beyond any alternative technology developed so far. Furthermore, graphene can be ideally employed to introduce intensity dependent losses into the external laser cavity.[30,31] To date, graphene has been only employed as external element, in transmission, to induce changes in the emission spectra of a THz QCL emitting over a limited (0.15 THz) spectral window.[32]

In the present work, we use a heterogeneous 17-μm-thick GaAs/AlGaAs heterostructure comprising three active modules, with gain bandwidths centred at 2.5 THz, 3 THz and 3.5 THz, respectively, and comparable threshold current densities ($J_{th}$). This leads to a very broad operational dynamic range of the heterogeneous gain medium: $J_{dr} = J_{max}/J_{th} = 3.2$, where $J_{max}$ is the maximum working current density.[15,17] Sample fabrication is based on a standard metal–metal processing technique, with lossy side absorbers lithographically implemented along the waveguide edges discussed in the Methods section.[15,33] The resulting lasers work as fully stabilized optical frequency comb synthesizers over 16% of their dynamic range,[15] as we recently demonstrated, by extracting its temporal profile and through the full-control and stabilization of the characteristic FCS parameters, i.e. the separation between adjacent modes and the carrier offset frequency, which were measured using a multi-heterodyne detection scheme in which the QCL-comb was mixed with a fully-stabilized optically-rectified THz-frequency comb.[24]

The graphene reflector is prepared by liquid phase exfoliation (LPE) of graphite in a water/surfactant solution (see Methods).[30] The resulting film is ~ 65 nm thick, as determined by atomic force microscopy[30] and covers a squared surface of about 1 cm$^2$. Raman spectroscopy is used to monitor the quality of the flakes at each step of the preparation process as well as to qualitatively estimate, in combination with electrical transport tests, the Fermi energy ($E_F \leq 250$ meV) of the flakes.[30] The resultant reflector (58% reflectivity in the THz)[30] behaves as an intraband driven (fast) saturable absorber at THz frequencies, providing 80% transparency modulation as a result of the intraband induced absorption bleaching.[30] This is confirmed when using our fully stabilized optical frequency comb in pulsed mode (10% duty cycle) (see Supplementary Information and Supplementary Figure S1).

The graphene reflector is then mounted on a copper mount via a thin indium foil (see Fig.1a) to ensure optimal thermal contact with the QCL copper mount with which it is in close thermal contact.

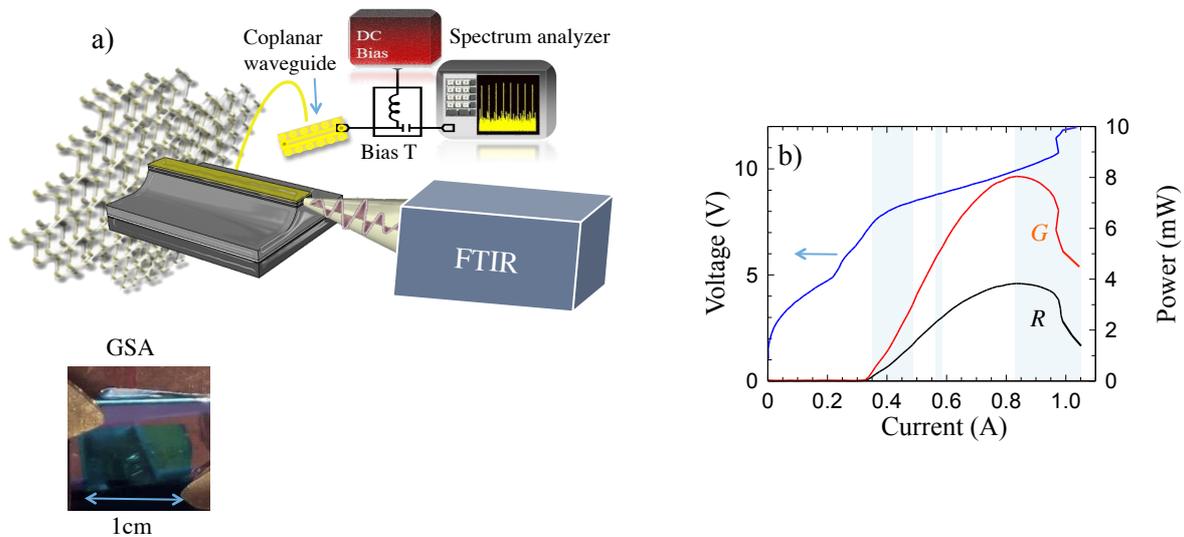

**Figure 1. THz QCL with an on-chip graphene saturable absorber (GSA) reflector**. **(a)** Schematics of a THz QCL comb coupled on chip with a GSA reflector; **(b)** Continuous-wave (CW) current-voltage (I-V) and light-current (L-I) characteristics of a 3.6 mm long, 50 µm wide THz QCL at a heat sink temperature of 15 K, without (*R*) and with (*G*) the on-chip GSA reflector. The optical power is measured with a calibrated power-meter (Thomas Keating Ltd) and corrected to take into account the 75% absorption of the cryostat window. The light blue shaded areas correspond to the regimes in which the laser behaves like an optical frequency comb synthesizer.

The graphene reflector is next tightly coupled on-chip, perpendicular to the back facet of a 3.6-mm-long, 50-µm-wide heterogeneous THz QCL, with a separation of 15 µm, meaning that it can be approximated as part of the laser cavity, and finely aligned by means of calibrated screws. A picture of the graphene saturable absorber reflector is shown in the bottom part of Fig. 1a. The unbiased reflector and the QCL, forming a coupled cavity with a 15µm air gap (Gires Tournois interferometer), were then mounted onto the cold finger of a continuous-flow helium cryostat. Light emission (Fig. 1b) is measured from the QCL using the facet opposite to the graphene reflector, as illustrated in the graphics of Figure 1a. This chip-scale embedded integration of the saturable absorber ensures long-term stability, appropriate thermal management, and accurate reproducibility of the experimental results.

In this configuration, the Gires Tournois interferometer does not affect the total group delay dispersion of the QCL, as shown by numerical simulations of the dispersion compensation performed using a finite element method (Comsol Multiphysics) (see Methods), shown in Figs 2a-2c.

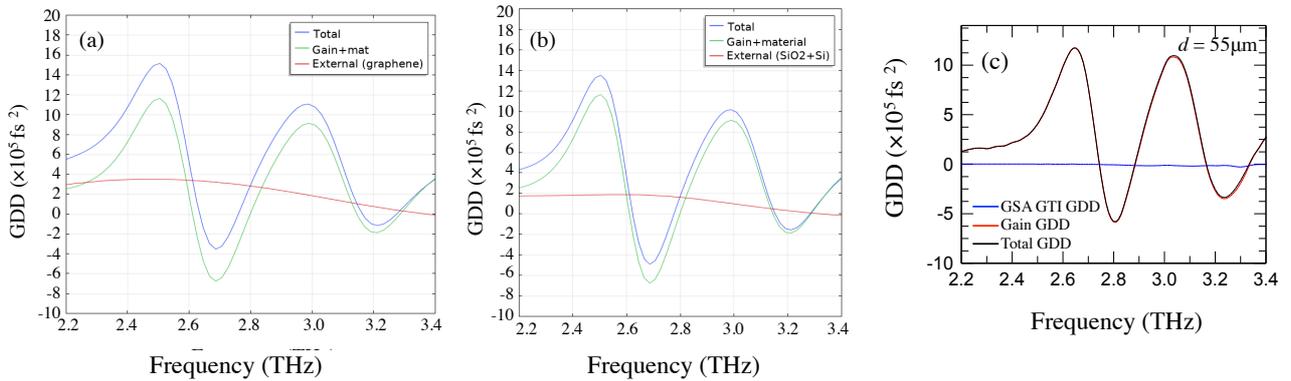

**Figure S2.** (a) Individual simulated group delay dispersion (GDD) of the GTI composed by the GSA reflector; the QCL comb material and gain GDD; and, the total GDD for the specific GTI length of 15μm. (b) Individual simulated GDD of the GTI composed of the 420 μm thick $SiO_2$/Si substrate; the QCL comb material and gain GDD; and, the total GDD for a GTI length of 15μm. (c) Individual simulated group delay dispersion (GDD) of the GTI composed by the GSA reflector; the QCL comb material and gain GDD; and, the total GDD for the specific GTI length of 55 μm, i.e for an "on-resonance GTI".

At a distance of 15 μm, the total group delay dispersion (GDD) does not show any significant variation, either when the QCL is integrated with the GSA reflector (Fig. 2a), or when it is combined with the bare 420 μm thick $SiO_2$/Si substrate (Fig. 2b). This is expected, as the phase change introduced by the GTI is almost negligible for such a small distance. Therefore, in this strong coupling condition, it is expected that dispersion compensation does not play any role in the phase-dynamics of the QCL comb modes, that are, in practice, unperturbed by the GTI, as in the case when a simple $SiO_2$/Si substrate is strongly coupled to the QCL. We then perform the same set of simulations while matching the distance $d$ with the first resonance of our GTI,[23] i.e. $d = 55$ $\mu$m (Fig. 2c). Very differently from what obtained with a coupled gold metal mirror [23] the GDD introduced by the GTI is more than one order of magnitude lower than that arising from the QCL gain, meaning that no dispersion compensation is expected to occur. We consider this effect to be a result of the high absorption of the graphene stack, which limits the feedback to the QCL.

In the following the graphene coupled-cavity laser is labeled as 'G', with 'R' referring to the same QCL with the on-chip graphene reflector.

Comparison between continuous-wave (CW) current-voltage (I-V) and light-current (L-I) characteristics (Figure 1b) of the graphene coupled cavity laser (sample G) and of the bare, QCL (sample R) shows that, whilst the device transport is not affected by passive coupled reflector, the optical power benefits from the non-linear reflector. An optical power of 8 mW is measured from the front facet of sample G, a factor of two larger than that of the bare sample R. Furthermore, the graphene reflector leads to factor of 2.3 increase in slope efficiency.

Considering the ± 40° divergence (Supplementary figure S2) measured from our double-metal THz QCL comb, and since the power emitted from the back facet of the QCL is initially equal to that measured from the reference sample $R$, it is expected that the intensity of the THz beam impinging on the graphene reflector varies from $I_o \sim 100$ W/cm$^2$ just above threshold (I = 400 mA) to $I_o \sim 800$ W/cm$^2$ at the optical peak power (I = 835 mA); this is well above the saturation intensity ($I_s \sim$ 6.3-6.7 W/cm$^2$) (Ref. 30 and supporting information) over the full dynamic range of the laser, confirming that the graphene film behaves as a saturable absorber.[30]

Representative CW Fourier transform infrared spectra under-vacuum (Figs. 3a-3d), measured for different currents in the *G* sample, show that well above threshold (>880 mA) (Figs 3b-3d) the GSA reflector does not induce major changes in the spectral behaviour of the laser (Fig. 3f-3h for comparison with *R* sample) over most of its operational range. It activates some additional in the 2.3-2.4 THz region (Fig. 3a-d), that persist over the whole operational range and which are absent in the bare laser (Fig.3e-3h).

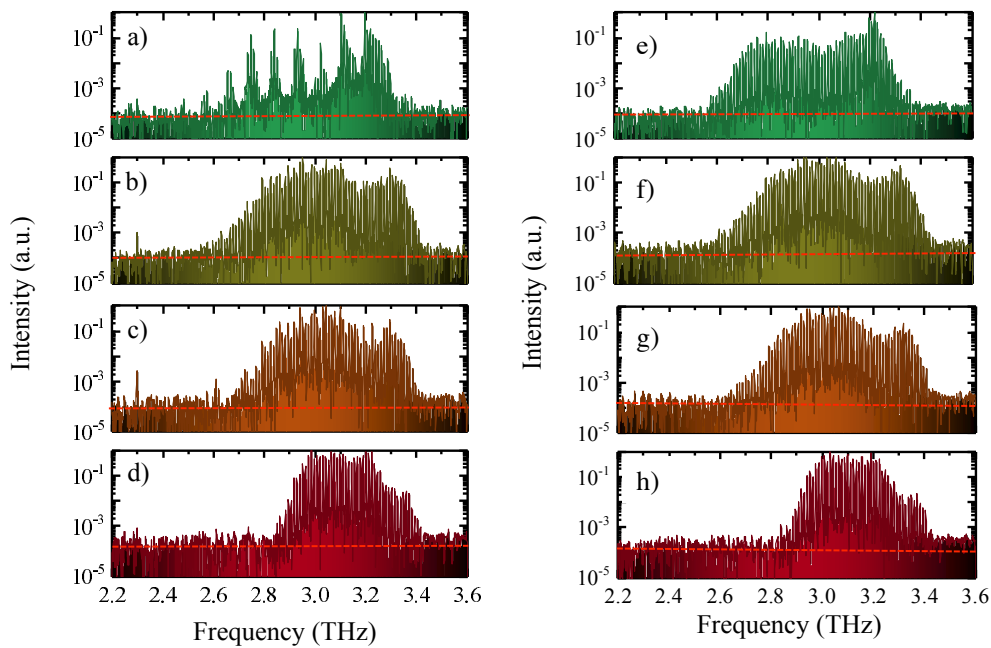

**Figure 3.** Vacuum Fourier transform infrared emission spectra. FTIR emission spectra collected under vacuum with a Bruker (Vertex 80) FTIR spectrometer in rapid scan mode, with a 0.075 cm-$^1$ resolution in the *G* sample (**a-d**) and in the *R* sample (**e-h**) The QCL was operated CW at a fixed heat sink temperature of 15K at driving currents of: (a,e) 560 mA, (b,f) 880 mA (c,g) 950 mA and (d,h) 1050 mA, respectively. The dashed horizontal lines indicate the noise floor of the measurements.

A remarkable difference is detected at 560 mA (J = 311 A/cm$^2$); the spectrum of the *G* sample in Fig. 3a has families of lasing modes individually spaced by the cavity round trip time, but the families are separated by a frequency matching a high order harmonic (9$^{th}$ harmonic) of the cavity's round trip frequency. Such a phenomenon is typical of harmonic mode locking, happening when multiple pulses per round trip are generated as a consequence of the modulation applied at harmonics of the cavity's fundamental round trip frequency.[34] The spectrum of the G sample then broadens gradually with bias reaching a continuous bandwidth of 0.94 THz (2.55-3.49 THz) at a current of 880 mA, and a discontinuous bandwidth of 1.25 THz (2.30 THz - 3.55 THz), with 8 mW of CW power, distributed amongst 90 equally spaced optical modes, as shown in Fig. 3b.

Figure 4a shows the corresponding free running electrical beatnote map, typically performed to characterize the coherence of the spectral emission of QCL-based frequency combs.[13]

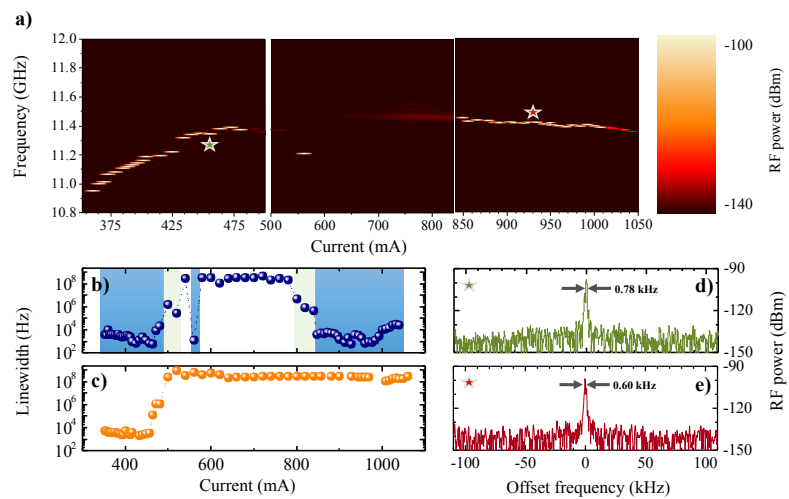

**Figure 4. Analysis of intermode beatnote. (a)** Intermode beatnote map as a function of QCL bias current for a 3.6 mm long, 50 μm wide laser bar integrated on-chip with a GSA reflector. The beatnote signal is extracted from the bias line with a bias-tee and it is recorded with an RF spectrum analyzer (Rohde and Schwarz FSW; RBW: 500 Hz, video bandwidth (VBW): 500 Hz, sweep time (SWT): 20 ms, RMS acquisition mode). All measurements are performed in CW at a fixed heat sink temperature of 15 K. The two starts two symbols indicate two representative driving current regimes (I=456mA) and (I=928 mA) in which the device behave like a stable frequency comb synthesizer. **(b-c)** Intermode beatnote linewidth as a function of driving current measured on (b) the GSA coupled QCL (*G* sample) and (c) the bare laser (*R* sample); **(d-e)** Intermode beatnote trace recorded at (c) I = 456 mA with center frequency at 11.345 GHz, and (d) at I = 928 mA with center frequency at 11.426 GHz, respectively.

Beatnote signals are extracted from the bias line, using a bias-tee, and recorded with a radio frequency (RF) spectrum analyser. A change in the intermode beatnote signal map is seen for the whole operational range of the QCL, when compared to that of the reference laser.[15,24] At $J = 228$ A/cm$^2$, when band alignment is fully reached, as typical of most QCL based frequency combs, the optical modes of the reference laser (sample R) lose their phase coherence and the intermode beatnote significantly broadens (>100 MHz linewidth) as the GVD becomes large enough to prevent FWM from locking the lasing modes in frequency and phase simultaneously.[15,24] This contrasts with the bias-dependent evolution of the beatnote map of the G sample (Fig. 4a). Specifically, the introduction of a multilayer GSA reflector causes a significant enlargement (from 16% (Fig 4c, and Ref. 15) to 55% of the laser operational range (Fig 4a) of the dynamic range in which THz QCL frequency comb is observed.

The comparative analysis of the beatnote linewidth (Figs. 4b-4c) clearly discloses the efficacy of the employed approach. Specifically, a set of individual beatnotes, as narrow as 780Hz (Fig. 4d), persists in the current range between 350 mA and 480 mA, as shown in Fig. 4b. The beatnote is a factor of 5 narrower than measured in the reference sample,[24] (Fig.4c), at specific biases/currents, thus suggesting that the GSA integration improves the phase locking of the optical modes. In analogy with the bare laser (sample R), [15,24] in the region between 350-480 mA, the RF beatnote signals are blue-shifted with a coefficient 3.6MHz/mA – this is a consequence of the changes in the relative distance between the beating modes, induced by the chromatic dispersion in the gain spectrum.

When the current in sample G is increased from 480 mA to 500 mA the single beatnote linewidth increases to 10 KHz, although it still preserves its narrow nature (Fig. 4b). Conversely, above 476 mA, the bare laser (sample R) (Fig. 4c)[15] looses its phase coherence[24] and develops a broad beatnote regime, reaching 300 MHz linewidth at 500 mA.

In the current range 500-525 mA the beatnote of sample G remains single, but superimposed to a broader beatnote (100 – 880 KHz). Over a very small (10 mA) current range, around 570 mA, the beatnote turns again single and narrow (950 Hz). This is a signature that the dispersion compensation in sample G enables phase locking of the emitted optical modes.

In the 580-780 mA range, the beatnote linewidth increases to 200-500 MHz, which is the signature of a transport regime dominated by dispersion. However, a single beatnote is still visible superimposed on a broader signal (see Fig.5a). Although wider, the presence of such an individual beatnote enables locking it to a microwave reference, in order to mode-lock the QCL and, consequently, reduce the beatnote linewidth.[35]

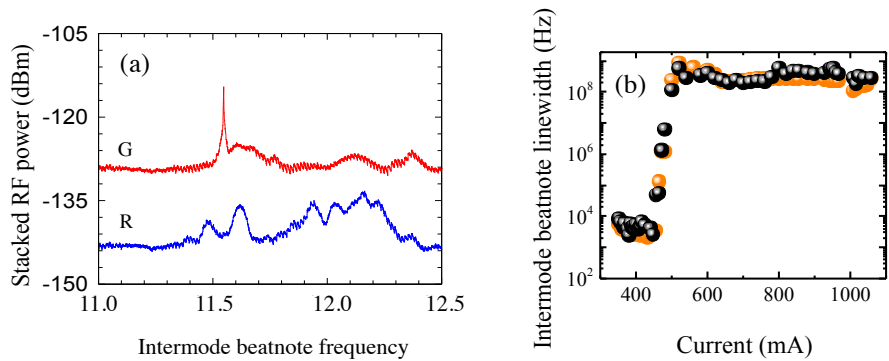

**Figure 5.** (a) Stacked intermode beatnote traces recorded at a heat sink temperature of 15 K whilst driving the GSA-coupled QCL (G) and the bare laser (R) at $J$ = 350 A/cm$^2$. The spectrum analyzer settings were: RWB: 10 kHz, VBW: 100 kHz, and SWT: 500 ms. The linewidth is limited by the RBW of the spectrum analyzer. (b) Intermode beatnote linewidth as a function of the driving current measured on the bare laser and while coupling it with the 420 μm thick SiO$_2$/Si substrate, placed at a distance of 15 μm from the QCL back facet.

Similarly, in the 780-835 mA, the beatnote remains single but turns narrower and is superimposed on a broader beatnote (100 – 790 KHz). In contrast, a fully dispersion dominated regime is observed in sample R over > 80% of its dynamic range (Fig. 4c),[15] i.e. at all current above 536 mA.

The most remarkable effect of the GSA reflector appears at driving currents > 845 mA (i.e. for $J$ = 2.6-3.2 $J_{th}$), where single beatnote linewidth preserves its narrow nature (600 Hz – 10 kHz) (Fig.4b and Fig. 4e). This corresponds to the high-phase noise regime that is typically recorded at larger current densities in the bare laser (sample R), and in all other THz QCL combs reported to date when $J$ > 1.16 $J_{th}$.

The observed single and narrow beatnote is in contrast to what retrieved in the R sample in which spontaneous FWM phenomena are typically unable to support dispersion compensation.[13,15,20] This is a unique characteristic of the GSA reflector with respect to any alternative saturable absorbers developed so far at THz frequencies such as n-doped semiconductors (GaAs, GaP and Ge). Although these materials can be used as THz SAs, they require electric fields of 10's kV/cm and cannot be easily integrated with a THz QCL using an approach similar to that described here, without inducing a major detrimental increase of the external cavity losses due to free carrier absorption.

By progressively driving the laser towards the negative differential resistance regime, the narrow beatnote red-shifts with a coefficient -0.33MHz/mA, mainly due to the local heating of the lattice at progressively higher driving currents.

Figures 4d and 4e plots intermode beatnote spectra in the two most interesting transport regimes in which in the QCL behave like a comb: 456 mA (Fig. 4d) and 928 mA (Fig. 4e). Beatnote linewidths as narrow as 780 Hz (Fig. 4d) and 600 Hz (Fig. 4e) can be retrieved, the narrowest values reported to date in any THz QCL FCS, to the best of our knowledge. Correspondingly, in the latter case, the spectrum shows a 0.94 THz spectral coverage, with a record output power (8 mW of CW power; 40 mW of peak power in pulsed mode, maximum wall plug efficiency 0.1%), distributed amongst 90 equally spaced optical modes. This represents state of the art in the field of THz QCL frequency combs. (It is worth noting that the retrieved linewidth values are ultimately limited by the jitter of the beat-note, since the laser is not stabilized.)

The physical explanation of the observed phenomenon at I > 845 mA can be found in the physical mechanism through which the graphene-related dynamics contributes to stabilize the QCL optical modes. Frequency and phase locking of the modes, which is the prerequisite of comb formation, can be obtained through four-wave mixing, which is generated by either fast saturable gain or loss in semiconductor lasers.[36] The former mechanism leads to a frequency modulated output, while the latter mechanism is associated with amplitude modulation. In THz QCL FC, both frequency and amplitude modulation are typically present,[37] and act simultaneously. In our setup, the gain and absorption are spatially separated, so they do not average out to a local net gain/loss, and create a spatially dependent profile within the cavity. The above arguments explain why, as a result of the interaction of the field emitted by the QCL with the inherently fast GSA,[30] and the related reinjection of this field into the laser cavity, the fast saturable loss of the graphene absorption layer contributes to the locking between the modes, which manifests itself by the observed extremely narrowed beatnote.

The dynamical processes behind the generation of such a sharp and narrow beat note, following the above arguments, are then investigated by performing time-domain simulations based on the Maxwell-Bloch equations in the configuration in which the QCL is uncoupled (sample R) or tightly coupled (sample G) with the GSA. We use the mbsolve[38] software, which is an open-source solver for the one-dimensional Maxwell-Bloch equations capable of handling spatial regions of different materials. We investigate the dynamical behavior of our experimental system, as described in the Supporting information file. The QCL is modeled as two-level gain medium, with the physical parameters summarized in Table I (supplementary material). Subsequently, we derive the intermode beatnote spectrum from the optical field, for both cases. The simulation results (Fig. S3, Supporting information), in agreement with our experimental findings, show a clear reduction of the intermode beatnote linewidth when the QCL is integrated with the GSA.

In addition to the saturable absorption effect inside the graphene, the reflector also features Fresnel reflection on the graphene surface. Since this component is spatially separated from the saturable absorption occurring inside the graphene, it might contribute to frequency comb stabilization through the same mechanism as the fast saturable gain in the active region. These effects then try to force the QCL into frequency-modulated operation, while the saturable absorption inside the graphene, helps regularize the remaining amplitude modulations.

The combination of the above effects can explain our experimental observations.

To further confirm our claims we perform a further set of experiments.

First, in Figure 5b we compare the reference unperturbed QCL beatnote evolution with the corresponding beatnote linewidth evolution recorded when the 420μm-thick Si/SiO$_2$ substrate is coupled (at 15 μm distance) to one end of the QCL laser cavity, exactly the same position of the GSA reflector. No difference is seen in the beatnote map with respect to the bare laser, confirming that the dispersion compensation, arising from the Si/SiO$_2$ 50% reflector, as predicted by numerical simulations (Figs. 2a-2b), does not play a role, and the GSA reflector is driving the inter-mode dynamics.

Secondly, we performed two set of experiments in which we collected the intermode beat-note linewidths by varying the distance of both the GSA and of the Si/SiO$_2$ substrate from the end of the QCL laser cavity ($d$), respectively, while keeping the current fixed at I = 900mA, i.e. in the region in which the most remarkable differences are unveiled.

In both cases, we first varied $d$ finely in the 15-25 $\mu$m range, then in the range 25-200 $\mu$m at which the impinging laser beam intensity $I \leq I_s$, and finally in the range 200-500 $\mu$m, where $I << I_s$. The rationale is that since GTIs are extremely sensitive to the distance from the laser facet, any eventual change of the beatnote map would be a signature of a GDD compensation operated by the GTI. The analysis of the intermode beatnote linewidth (Figures 6a-6b) at 900 mA, unambiguously shows that in our G sample, by varying $d$ in and out resonance, the intermode beatnote remains single and preserves its narrow nature (Fig. 6a) for $d \leq 200$ $\mu$m unlike what experimentally observed in the case of a dispersion compensated GTI THz QCL comb,[23] achieved by coupling a metal mirror.

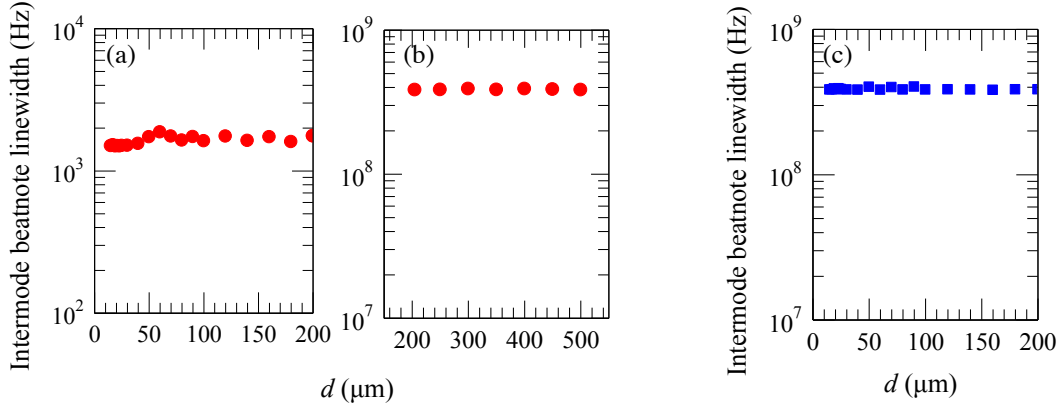

**Figure 6.** (a-b) Evolution of the electrical intermode beatnote linewidth of the GSA-coupled QCL as a function of the distance d of the GSA from the laser facet, for a fixed driving current of 900 mA. (c) Evolution of the electrical intermode beatnote linewidth of the QCL coupled with a 420-$\mu$m-thick $SiO_2$/Si substrate, as function of the distance d of the $SiO_2$/Si substrate from the laser facet, for a fixed driving current of 900 mA. All measurements were performed in CW at 15 K. The beat-note signal is extracted from a bias line, employing a bias-tee connected to a RF optical spectrum analyzer (OSA) in RMS acquisition mode (OSA setting: RWB: 500 Hz, VBW: 500 Hz, SWT: 50 ms)

In this latter case, a narrow (~3 kHz) and intense beatnote is only achieved at periodic positions of the external metal coupling element.[23] For $d > 205$ $\mu$m (Fig. 6b), the GSA-integrated QCL behavior is identical to that of the bare laser, which is a signature that the GSA plays no role. As expected, in the Si/$SiO_2$ – coupled QCL the distance does not affect the broad linewidth (> 100 MHz) that persists over the whole spanned range of distances (Fig. 6c).

The ability to coherently frequency stabilize of the GSA reflector is further proved by applying a direct RF modulation at its round-trip frequency. Injection of a periodic signal into an oscillator is commonly used to stabilize emission frequency and/or laser cavity modes separation,[39] and corresponds to round-trip frequency stabilization as normally achieved by injection of an external microwave signal on the driving current. Ref. 35 demonstrated injection locking of the intermode frequency difference of THz QCL over hundreds of MHz by driving the laser bias with a microwave signal close to this frequency. Ref. 40 achieved the stabilization of the frequency difference between two lateral modes of a THz QCL with corresponding frequency locked linewidths ≤ 10 Hz and with a negligible drift. Here, we demonstrate the stability of our GSA-coupled QCL-comb inter mode separation by measuring the shape of its photocurrent spectrum in the microwave range, around the cavity round-trip frequency, where a sharp single peak proves stable phase and frequency relation between adjacent laser modes.

Figs. 7a-7b show all-electrical injection locking of a narrow beatnote to the RF oscillator and the corresponding locking range as function of the RF-power transmitted inside the cavity of sample *G*; this follows the square root behaviour predicted by Adler's equation.[35]

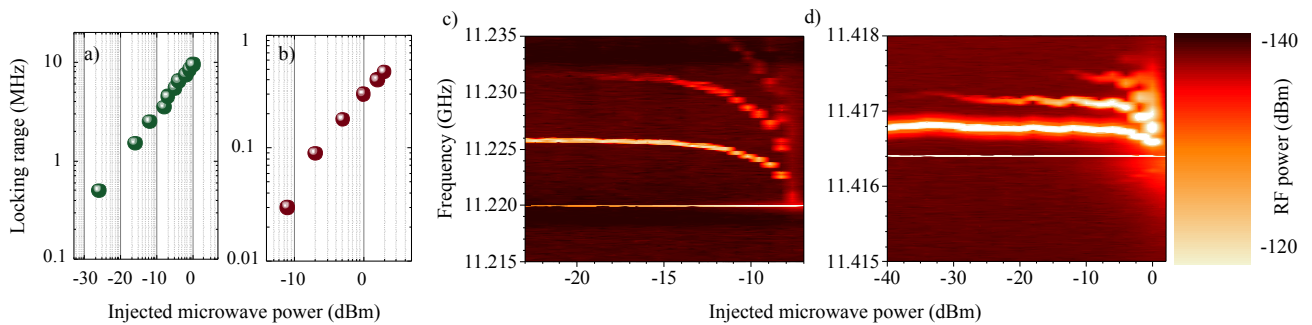

**Figure 7. Coherent injection locking. (a-b)** Locking range as a function of RF-power transmitted inside the cavity of a THz QCL-comb integrated on-chip with a GSA reflector. The RF and dc-bias signals are brought to the QCL using a 60GHz, 50 ohm coplanar probe positioned at one end of the QCL ridge. The intermodal beatnote (IBN) frequency is at a repetition frequency of ~11 GHz, and is electrically extracted from the QCL by connecting the RF-synthesizer to the input port of a 40 GHz directional coupler, with the output port of the coupler connected to the ac input of a 40 GHz bias-T. The dc line of the bias-T is connected to the power supply that drives the QCL at a constant current of (a) 425 mA, and (b) 980 mA. The spectra are collected with the same settings of Fig. 3a. **(c-d)** IBN spectra for different RF injected powers. The RF frequency is held constant at approximately (c) 6 MHz or (d) 0.4 MHz apart from the cavity round trip frequency and the laser is operating at (c) 425 mA, and (d) 980mA. Microwave spectra are on a logarithmic scale for different injected powers.

The RF-signals capture the comb oscillator, imposing an oscillation frequency even when the RF power is as low as -29dBm (Fig. 7a, I = 400 mA), provided that the two oscillators are at sufficiently close frequencies and their coupling is strong enough. Equally, with a higher laser current (Fig. 7b, 980 mA), locking is observed, albeit with higher injection power levels.[39]

To investigate the injection locking dynamics further, we retrieve the beat note spectra, measured in the single beatnote regimes, as the injected RF power is increased. The broad beatnote is pulled towards the frequency of the injected signal (11.2200, Fig.7c; 11.4164, Fig. 7d) and finally locked, as shown in Figs 4c-d at the different bias currents of 425mA (Fig. 7c) and 980mA (Fig. 7d), respectively. Two side peaks that are ~ 9 dBm weaker than the initial beatnote remain for RF injected powers up to -8 dBm in Fig. 7c, then vanish, and the microwave spectrum of the intermode beating is fully controlled by the injected signal. The noise floor around the locked narrow beatnote is ~ 20 dB, i.e. weaker than the peak power of the originally broad beatnote. This proves that the intermode beatnote power is almost completely locked.

In conclusion, we demonstrated that the free-running phase coherence of broadband THz QCLs with a heterogeneous active region can be substantially enhanced by on-chip coupling to a saturable absorber prepared from a liquid phase exfoliated graphite film. Indeed, with the on-chip integration of a GSA reflector on a fully stabilized THz comb, operating only over 16% of its dynamic range,[24,15] we achieve stable comb operation over >55% of the laser operational range, with a beatnote linewidth as narrow as 600 Hz, 8mW of CW power and over 90 equally spaced optical modes covering a 0.94 THz spectral bandwidth, with over three-decade reduction of phase-noise over an additional 15% of this range. We also achieve injection locking over the same operational range as the FCS, showing the stability of the comb operation. Our compact and frequency agile design, together with the high optical power, the extremely large number of optical modes and the narrow beatnote linewidth, well suitable for dual-comb THz spectroscopy[41] and hyperspectral imaging[42] opens up opportunities to deliver an all-in-one miniaturized nano-engineered frequency comb electrical source for spectroscopic and nanoscale applications in the far infrared. Future perspectives of the present work include the experimental assessment of the potential of the present design, in conjunction with graphene's fast recovery time and saturation fluence, to produce time domain pulses with sub-ps widths and average power comparable to the CW operation level of a QCL.

**Methods**

**QCL fabrication**

Lasers are fabricated in a metal–metal waveguide configuration. We first perform Au-Au thermo-compression wafer bonding of the 17-μm-thick active region (sample L1458) onto a highly doped GaAs substrate, followed by the removal, through a combination of mechanical lapping and wet etching, of the host GaAs substrate of the molecular beam epitaxial (MBE) material. The $Al_{0.5}Ga_{0.5}As$ etch stop layer is then removed using HF etching. Laser bars are then defined by inductively-coupled plasma etching, which leads to almost vertical sidewalls (hence uniform current injection into the full gain region). Following etching, a Cr/Au (10 μm/150 μm) top contact is deposited along the center of the ridge surface, leaving a thin region uncovered along the ridge edges. 5μm-wide Ni (5-nm-thick) side absorbers were then deposited over the uncovered region using a combination of optical lithography and thermal evaporation. Such a lossy side absorbers are intended to inhibit lasing of the higher order lateral modes by increasing their threshold gain.[5] Finally, the backside of the substrate is lapped down to 150 μm to optimised thermal management and enable operation in CW. Laser bars, 50 μm wide and 3.6 mm long, are then cleaved and mounted on a gold coated copper bar, and then onto the cold finger of a He continuous-flow cryostat.

**Preparation of the graphene reflector**

The water-based ink is prepared by ultrasonicating (Fisherbrand FB15069, Max power 800W) graphite flakes (Sigma Aldrich) for 9 h in deionized water with sodium deoxycholate (SDC, 9 mg ml$^{-1}$). This is then vacuum filtered using 100 nm pore-size nitrocellulose filters. This blocks the flakes, while water passes through, leading to a film on the surface of the filter. This is then placed on a 420μm-thick intrinsic high-resistivity Si/SiO$_2$ double polished wafer (acting as a back reflection mirror) and annealed at ~80 °C for 2 h, to improve adhesion, followed by dissolution of the filter in acetone overnight.

**Simulation of the total group delay dispersion of the *G* sample**

Numerical simulations of the group delay dispersion (GDD) (Figs. S2(a) and S2(b)) are performed using a finite element method (Comsol Multiphysics). The simulated structure includes the end of the QCL waveguide and a 65 nm thick GSA on a 420 μm thick SiO$_2$/Si mirror and separated by 15 μm from the laser back facet; this structure was then surrounded at a distance of a few λ by vacuum and absorbing boundary conditions. As the SiO$_2$ layer is very thick (~300μm) compared to the separation between the graphene and the QCL facet, the amount of radiation that could be reflected back to the QCL from the SiO$_2$/Si boundary is expected to be negligible. For such a reason, absorbing boundary conditions are set at about 100μm inside the SiO$_2$, as well as at all the other external boundaries of the simulation domain. The imaginary part of the refractive index of the graphene is computed from the experimental values of the absorption coefficients reported in ref 1 while the real part is numerically computed considering a graphene layer having an electron temperature of 20 K, a total scattering time τ = 0.1 ps and a chemical potential μ$_F$ of 250 meV. The variation of the imaginary part of the graphene under illumination from the QCL is computed according to the expected reduction of the saturable absorption coefficient. The variation of the real part is then computed applying the Kramers-Krönig equations.

THz radiation was injected into the QCL waveguide (from the end opposite to the GTI) and then reflected back into the QCL waveguide by the GTI. This allows us to obtain the amplitude and phase of the S$_{11}$ scattering parameter.[22,23] Finally, the dispersion provided by the structure was computed from the second derivative of the phase. The final GDD profile took into account the contributions from the semiconductor material and gain of the QCL,[23] and that of the GTI. The frequency dependent refractive index of the material is computed using a Drude-Lorentz model, adding its deviation due to the QCL's gain, obtained from the Kramers-Kronig equations. The waveguide dispersion contribution, negligible with respect to the other terms, was neglected

**All-electrical injection locking of a narrow beatnote to an RF oscillator**

The modulation signal was supplied by an external stabilized RF synthesizer (Rohde & Schwarz SMA100B) through a Bias-T (Tektronix PSPL5544). The beatnote signal was then extracted from the bias line using the same bias-tee, and recorded with an RF spectrum analyser (Rohde & Schwarz FSW). For this, a high-speed Sub-Miniature Push-on (SMP) connector with an integrated coplanar transmission line waveguide is used, allowing high-frequency electrical signals to pass for fast electrical modulation, and suppression of losses and deformation at the wire bonding points. The electrical connections between one end of the QCL ridge and the coplanar probe are realized using short and thin (20 μm) Au bonding wires.

**Supporting Information**

This material is available free of charge via the internet at http://pubs.acs.org.

**Acknowledgements**

The authors acknowledge financial support from the ERC Project 681379 (SPRINT), the EU union Graphene Flagship (core 3 project) Hetero2D, GSYNCOR, MINERGRACE; EPSRC Grants EP/K01711X/1, EP/K017144/1, EP/N010345/1, and EP/L016087/1; and the EU Graphene and Quantum Flagships. EHL acknowledges support from the Royal Society and the Wolfson Foundation.

**Competing financial interests**: The authors declare no competing financial interests.

**Data availability**: The data that support the plots within this paper and other findings of this study are available from the corresponding authors upon reasonable request.

**Additional information**

Supplementary Information is linked to the online version of the paper

**Table of Contents (TOC) Graphic**

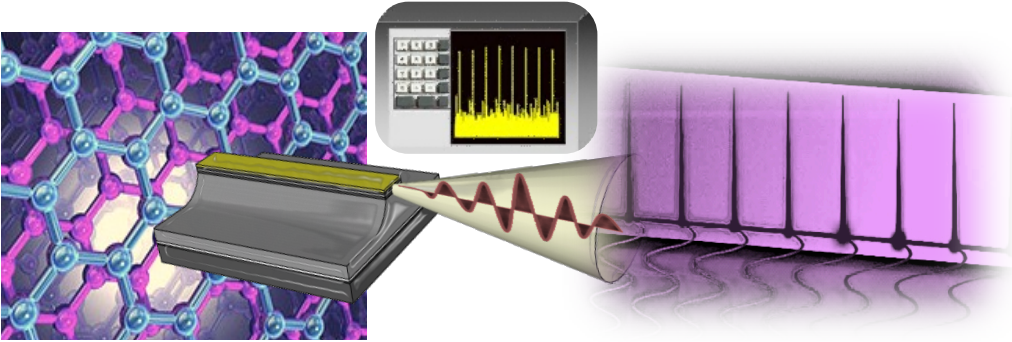

# Terahertz frequency combs exploiting an on-chip solution processed graphene-quantum cascade laser coupled-cavity architecture

Francesco P. Mezzapesa,[1] Katia Garrasi,[1] Johannes Schmidt,[1] Luca Salemi,[1] Valentino Pistore,[1] Lianhe Li,[2] A. Giles Davies,[2] Edmund H. Linfield,[2] Michael Riesch,[3] Christian Jirauschek,[3] Tian Carey,[4] Felice Torrisi,[4+] Andrea C. Ferrari,[4] and Miriam S. Vitiello[1*]

[1] *NEST, CNR - Istituto Nanoscienze and Scuola Normale Superiore, Piazza San Silvestro 12, 56127, Pisa, Italy*

[2] *School of Electronic and Electrical Engineering, University of Leeds, Leeds LS2 9JT, UK*

[3] *Department of Electrical and Computer Engineering, Technical University of Munich, Arcisstr. 21, 80333 Munich, DE*

[4] *Cambridge Graphene Centre, University of Cambridge, Cambridge, CB3 0FA, UK*

[+] *present address: Molecular Science Research Hub, Imperial College London, London W12 0BZ, UK*

[*] miriam.vitiello@sns.it

**Open Aperture z-scan**

To investigate the THz-induced non-linear absorption properties of the GSA reflector, we use an open-aperture z-scan technique.[1] The THz radiation generated by our QCL comb, driven in pulsed mode (10% duty cycle, pulse width 1μs, I = 450 mA) is focused onto the graphene sample at normal incidence using two closely positioned convergent lenses with 3 cm focal length (Ref. 29). The GSA is then placed on a micrometric translation stage, which allowed its position to be varied along the focal axis. A pyroelectric sensor is then positioned at a fixed distance from the laser facet, behind the sample holder, to collect the transmitted radiation. The same experiment is repeated on the uncoated substrate to normalize the transmittance data.

Figure S1 plots the z-scan transmission data. The results are consistent with Ref. 1, achieved by employing a 3.4 THz single-plasmon QCL, with a transmission enhancement of 80%, the non-saturable and saturable components of the linear absorption $\alpha_{NS} = 0.15\pm0.02$ and $\alpha_{S} = 0.70\pm0.025$, respectively, and saturable intensity $I_S = 6.3\pm1.3$ Wcm$^{-2}$, well below the intensity impinging on the GSA over the entire QCL dynamic range.

In the experiment described in the main text, the radiation emitted from the back QCL facet first impinges onto the graphene layer, then it is partially reflected (50%) and partially transmitted (50%) by the first facet of the SiO$_2$/Si substrate and again partially transmitted (50%) and partially reflected (50%) by the second facet of the SiO$_2$/Si substrate. While the final transmitted beam portion is that retrieved by the pyroelectric sensor in the z-scan transmission experiment, the final reflected beam experiences a further 50% reflection and 50% transmission at the first SiO$_2$/Si substrate facet before impinging again on the GSA. This means that the overall power impinging on the GSA before light is recoupled into the laser cavity is significantly larger than $I_s$ and the GSA is almost transparent in this final round trip.

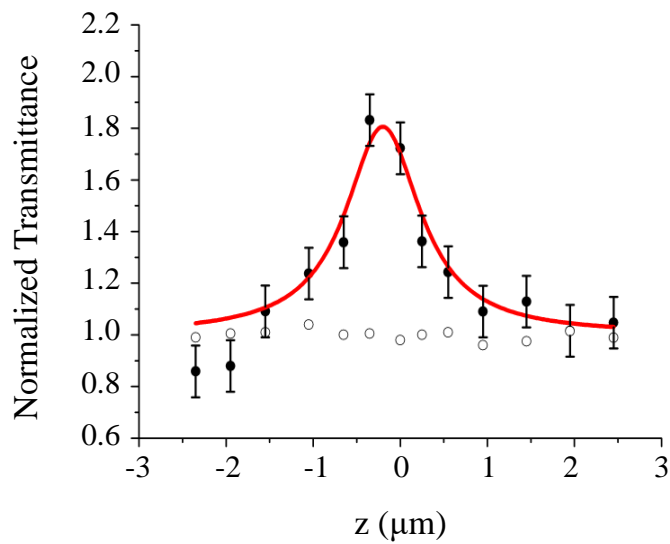

**Figure S1**. z-scan normalized transmittance traces of the GSA (full dots) probed with a QCL frequency comb operating at 10% duty cycle, compared with the z-scan normalized transmittance traces of the Si/SiO$_2$ substrate (empty dots). The error bars correspond to the uncertainty interval on the measured normalized transmittance. The dashed lines are fits of the normalized transmittance T(z) using the following equation (Ref. 1): $T(z) = \left[1 - \alpha_0 + \alpha_S - \frac{\alpha_S\left(1+\frac{z^2}{z_R^2}\right)}{1+\frac{z^2}{z_R^2}+\frac{I_0}{I_S}}\right]\frac{1}{1-\alpha_0}$ where $I_0$ is the beam intensity at the focal point, $z_R$ is the Rayleigh length, $\alpha_{NS}$ and $\alpha_S$ represent the non-saturable and saturable components of the linear absorption $\alpha_0 = \alpha(I = 0) = \alpha_{NS} + \alpha_S$, respectively, and $I_S$ is the saturation intensity.

**Far-field profile of the THz QCL comb**

The far-field intensity profile of the double metal THz QCL is measured by scanning a pyro-electric detector, with a sensitive detection area of 7mm$^2$, using a motorized stage to scan a spherical surface. The far-field pattern shown in Figure S2 is measured at a heat sink temperature of 15 K with the QCL being driven with a current I = 800 mA.

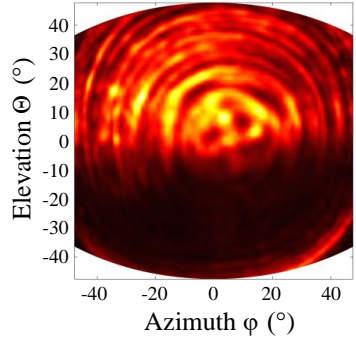

**Figure S2.** Far-field intensity profile of the double metal THz QCL comb measured whilst driving the QCL with a current I = 800mA at 15 K, and placing the pyroelectric detector at 6 cm from the laser facet.

**Time domain simulation of the intermode beat note linewidth**

In order to provide theoretical insights into the dynamical processes behind the generation of a single and narrow intermode beat note at high driving currents (I > 845mA), we perform numerical simulations based on the Maxwell-Bloch equations.[2] These equations are used routinely for time-domain simulations of QCLs[3] and can also allow the GSA reflector to be modelled properly. Here, we use the full-wave Maxwell-Bloch equations, i.e., without invoking the rotating wave approximation. These inherently include spatial hole burning, which plays an important role in the QCL frequency comb formation process.[4] The simulations are started from noise, i.e., the electric field is initialized with random numbers from a Gaussian distribution, and the magnetic field is initially set to zero. The QCL active region is modeled as two-level gain medium, with the physical parameters summarized in Table I, appropriately chosen from a set of available values,[7,8,2] to match the measured experimental laser spectrum. For graphene, the experimentally measured intensity dependent loss coefficient is used. Saturable absorption is treated as a limiting case of the Bloch equations, as detailed below.

In a GSA, the intensity loss coefficient can be written as:[1]

$$\alpha(I) = \alpha_{NS} + \frac{\alpha_S}{1 + (I/I_S)} \quad (1)$$

depends on the light intensity I. Here, $I_S = 63$ kW m$^{-2}$ is the saturation intensity and $\alpha_S = 0.70$ and $\alpha_{NS} = 0.15$ are the saturable and non-saturable loss coefficients, respectively, measured via the open aperture z-scan experiments. The intensity-dependent loss term $\alpha(I)$ denotes the power loss of our 65 nm thick graphene region. In order to convert the non-saturable part $\alpha_{NS}$ to a linear amplitude loss coefficient (as used in mbsolve), we divide the value by 2 x 65 nm. The factor 2 is not required for the saturable part, which is modeled by the two-level system, since Eq. (2) assumes that $\alpha_S$ is a power loss coefficient. Saturable absorbers can be generally described as two-level systems[2,7,8] using the relations

$$\alpha_S = \Gamma \omega_{21} N_{3D} \frac{|d_{21}|^2 T_2}{n_{eff}} \frac{1}{\varepsilon_0 c \hbar} \quad (2)$$

and

$$I_S = \hbar^2 \varepsilon_0 c \frac{n_{eff}}{|d_{21}|^2 T_2} \frac{1}{2T_1} \quad (3)$$

We employ the adiabatic elimination and assume instantaneous absorption (i.e., T1, T2 ≈ 0). For convenience, we use the values from the QCL model for the background refractive index, the overlap factor, and the transition frequency. Then, we can determine the values for the dipole moment $d_{21}$ and charge carrier density $N_{3D}$:

$$|d_{21}|^2 = \hbar^2 \varepsilon_0 c \frac{n_{eff}}{I_S T_2} \frac{1}{2T_1} = 1480\ nm \quad (4)$$

and

$$N_{3D} = \frac{2 \alpha_S T_1 I_S}{\Gamma \hbar \omega_{21}} = 6.43 \times 10^{20}\ m^{-3} \quad (5)$$

using the parameters listed in Table 1.

The non-saturable part of the loss term is considered as background loss and set accordingly during construction of the material. We modeled our G sample without including the 15 µm air gap, for simplicity. As its size is significantly smaller than the central wavelength (103 µm), its effect on the operation is limited to loss, which is independent of frequency and intensity, and so can be included elsewhere (e.g., in the QCL region). The GSA can be then considered as part of the laser cavity.

Similarly, the outcoupling of radiation through the QCL facets is incorporated in our model resulting in a slight increase of the waveguide losses. This enables simple boundary conditions, as, in this case, both QCL facets as well as the GSA substrate can be assumed to be perfectly reflecting. Therefore, the modeled experimental configuration only consists of the QCL, in the reference case (bare QCL), and of the same configuration at which we simply added the GSA for our GSA-coupled QCL (sample G).

The ultrathin thickness (65 nm) of the GSA constitutes a major challenge for common finite-difference time-domain (FDTD) solvers, as the spatial discretization size must be reduced significantly. In order to maintain computational efficiency, we use a further approximation. We assume that the main effect of the graphene is the reflection at the interface, and that the radiation that enters the absorber is rapidly attenuated. Then, we can artificially extend the dimension of the absorber.

| Name | Symbol | QCL (gain) | Graphene (absorber) |
|---|---|---|---|
| Charge carrier density | $N_{3D}$ | $4 \times 10^{21}$ m$^{-3}$ | $6.43 \times 10^{20}$ m$^{-3}$ |
| Transition frequency | $\omega_{21}$ | $2\pi \times 3$ THz | $2\pi \times 3$ THz |
| Dipole moment | $d_{21}$ | 6 nm | 1480 nm |
| Recovery time | $T_1$ | 10 ps | 0.15 ps |
| Dephasing time | $T_2$ | 555 fs | 100 fs |
| Equilibrium inversion | $\Delta_{eq}$ | 1 | -1 |
| Amplitude loss coefficient | $\alpha_{NS}$ | 8 cm$^{-1}$ | $1.15 \times 10^4$ cm$^{-1}$ |
| Refractive index | n | 3.6 | 3.6 |
| Overlap factor | $\Gamma$ | 1.0 | 1.0 |
| Length | L | 3.6 mm | 5 µm |

**Table I.** List of physical parameters and corresponding values used to describe the QCL and graphene domains in the simulation performed with the mbsolve[9] software.

Table I summarizes all simulation parameters of both QCL and GSA. Both experimental conditions (bare QCL, and GSA coupled QCL) are simulated using 2048 spatial grid points, which roughly corresponds to $\lambda/10$, where $\lambda$ is the maximum wavelength of the QCLs spectrum. We record the electric field at the front facet of the QCL with a sampling time of 29 fs, which corresponds to 10 points per maximum time period. The simulation end time is set to 1.5 µs, which yields a spectral resolution of about 667 kHz (Fig. S3(b). While the spectral resolution of the experiment cannot be achieved due to unrealistic simulation runtimes, well beyond our supercomputer capability, it is sufficient to well capture the difference in the intermode beatnote linewidth as depicted in Fig. S3 (a). The intermode beat note is determined by calculating the power at the microwave frequency $f_{GHz}$ to be proportional to the spectral intensity products of all possible points in the THz spectrum, separated by $f_{GHz}$. The simulation results well reproduce the strong reduction of the intermode beatnote linewidth operated by the integrated GSA.

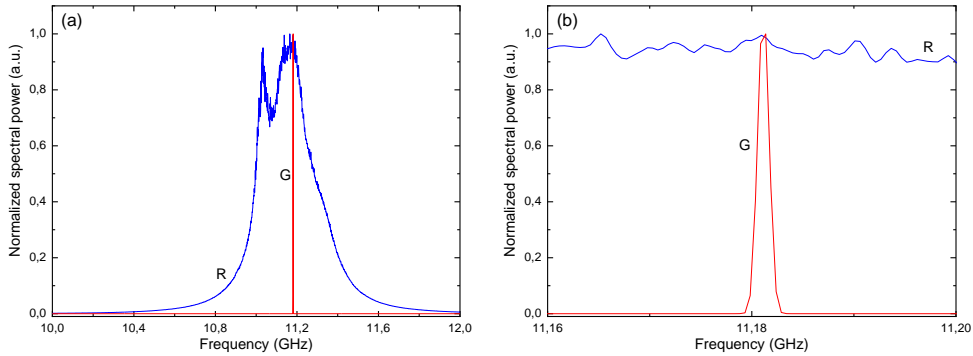

**Figure S3.** (a) Simulated beatnote power spectrum for the bare laser (blue trace) (sample *R*) and for the device with the GSA (red trace) (sample *G*). The parameters employed to describe the simulated systems are summarized in Table I. (b) Zoom on the intermode betanote of sample *G*. A further reduction of the intermode beatnote linewidth is expected for increased simulation times, beyond the actual capabilities of the employed computer.